\documentclass{article}

\usepackage{arxiv}

\usepackage[utf8]{inputenc} 
\usepackage[T1]{fontenc}    
\usepackage{url}            
\usepackage{booktabs}       
\usepackage{amsfonts}       
\usepackage{nicefrac}       
\usepackage{microtype}      
\usepackage{lipsum}

\usepackage{lmodern}
\usepackage{graphicx}
\usepackage[figurename=Fig.,labelfont=bf,labelsep=period]{caption}
\usepackage{subcaption}
\usepackage{amsmath}

\usepackage{newtxtext,newtxmath}
\usepackage[colorlinks=true,citecolor=red,linkcolor=black]{hyperref}

\newcount\Comments  
\usepackage{color}
\definecolor{darkgreen}{rgb}{0,0.5,0}
\definecolor{purple}{rgb}{1,0,1}
\newcommand{\kibitz}[2]{\ifnum\Comments=0\textcolor{#1}{#2}\fi}

\newcommand{\edit}[1]{\kibitz{black}      {#1}}

\title{Calibrating microscopic car following models for adaptive cruise control vehicles: a multi-objective approach}

\author{
  Felipe de Souza\\
  Postdoctoral Appointee\\
  Division of Energy Systems\\
  Argonne National Laboratory\\
  Lemont, IL, USA\\
  \texttt{fdesouza@anl.gov} \\
   \And
  Raphael Stern\\
  Assistant Professor\\
  Department of Civil, Environmental, \\and Geo- Engineering\\
  University of Minnesota\\
  Minneapolis, MN, USA\\ 
  \texttt{rstern@umn.edu} \\
}

\begin{document}
\maketitle

\begin{abstract}
\textit{Adaptive cruise control} (ACC) vehicles are the first step toward comprehensive vehicle automation. However, the impacts of such vehicles on the underlying traffic flow are not yet clear. Therefore, it is of interest to accurately model vehicle-level dynamics of commercially available ACC vehicles so that they may be used in further modeling efforts to quantify the impact of commercially available ACC vehicles on traffic flow. Importantly, not only model selection but also the calibration approach and error metric used for calibration are critical to accurately model ACC vehicle behavior. In this work, we explore the question of how to calibrate car following models to describe ACC vehicle dynamics. Specifically, we apply a multi-objective calibration approach to understand the tradeoff between calibrating model parameters to minimize speed error vs. spacing error. Three different car-following models are calibrated for data from six vehicles. The results are in line with recent literature and verify that targeting a low spacing error does not compromise the speed accuracy whether the opposite is not true for modeling ACC vehicle dynamics.
\end{abstract}


\section{Introduction}\label{sec1}The impact that \textit{autonomous vehicles} (AVs) will have on traffic behavior and dynamics has been an area of significant research activity in the past years~\cite{talebpour2016influence,stern2017dissipation,Cui2017,ge2018experimental,darbha1999,bose2003analysis,davis2004effect,kesting2010enhanced,shladover2012impacts,van2006impact,delis2016simulation,dollar2018efficient}. However, before vehicles reach full automation there will be a transition period in which vehicles with partial automation will become more prevalent on our roadways~\cite{bansal2017forecasting}. For example, technologies such as \textit{adaptive cruise control} (ACC) capture much of the longitudinal control of AVs and are commercially available on many new vehicles being sold today. It has been shown that the behavior of only a relatively small number of AVs or ACC vehicles can substantially impact the traffic flow ~\cite{talebpour2016influence,stern2017dissipation,davis2013effects}. Therefore, there is significant interest in understanding the the dynamics of ACC vehicles which can provide a clear picture of what the impact of ACC vehicles on traffic flow, stability, and throughput.

An essential first step in modeling the impact of ACC vehicles is to obtain accurate models that describe ACC vehicle behavior under different traffic conditions. These models can then be used in simulation to analyze how increased vehicle automation will affect traffic flow and throughput ~\cite{davis2004effect,davis2013effects,talebpour2016influence} as well as emergent properties such as string stability, defined as whether small perturbations are not amplified by the follower vehicles~\cite{wilson2011car}. 

Much work has been done to design ACC controllers that have favorable traffic flow properties such as a small following headway (time gap) while maintaining string stability~\cite{levine1966optimal,ioannou1993intelligent,shladover1995review,swaroop1996string,rajamani1998design,liang1999optimal,alam2015heavy,orosz2005bifurcations,besselink2017string,monteil2018l2}. However, recent evidence has indicated that the ACC controllers that are being commercially implemented and available on consumer vehicles may not be string stable~\cite{milanes2014modeling,gunter2019model,gunter2019are}. This can lead to further consequences such as increased energy consumption~\cite{stern2017dissipation} and emissions~\cite{stern2018emissions} due to unnecessary speed oscillations. The overall impacts on the traffic flow of commercially available ACC vehicles are still unclear as data are still limited and models are still not able to capture ACC behavior in all its nuances. Therefore, it is necessary to develop and analyze microscopic traffic flow models that describe the dynamics of commercially available ACC vehicles with higher accuracy. These should be constructed using experimentally collected data that captures the ACC car following behavior.

The question of how best to model the dynamics of an individual vehicle as a function of the behavior of the vehicles around  has been of interest since the early car following experiments that were conducted in the 1950s~\cite{gipps1981behavioural,chandler1958traffic,herman1959single,rothery1964analysis,gazis1959car}. For a vehicle at position $x$, the general framework of second-order car following models is to describe the acceleration of the following vehicle $\ddot{x}$ as a function of the inter vehicle spacing $s$, the speed of the following vehicle $\dot{x}$, and the relative velocity between the lead vehicle and the following vehicle $\dot{s}$:
\begin{equation}
    \ddot{x} = f(s, \dot{x}, \dot{s}),
\end{equation}
\noindent where each successive derivation is indicated with an additional dot. While not all car following models fit into this general form (the Gipps model being a notable exception since it is inherently modeled in discrete time), this is a general form that many models take.

In the years since the early car-following experiments, many microscopic traffic flow models have been proposed and validated including the Gipps model~\cite{gipps1981behavioural}, Bando's \textit{Optimal Velocity model} (\edit{OVM})~\cite{BandoHesebeNakayama1995}, the Gazis-Herman-Rothery (GHR) model~\cite{chandler1958traffic}, \edit{the \textit{Optimal Velocity Relative Velocity model} (OVRV)~\cite{milanes2014modeling},} and the \textit{Intelligent Driver model} (IDM)~\cite{treiber2000congested}, among others. One important characteristic of car-following models is whether they explicitly model delays (reaction times for drivers or sensor/actuation delay in the case of automated vehicles), or they are delay free models. Both approaches have been show to be able to accurately capture macroscopic properties of traffic flow dynamics~\cite{BandoHesebeNakayama1995,chandler1958traffic}. Most calibration efforts with these models have been to find the best model parameter values to describe human car following behavior. However, these models can also be applied to the vehicle dynamics of partially autonomous vehicles as well~\cite{gunter2019are}. In fact, using car following models to describe the behavior of ACC vehicles may be particularly successful since ACC vehicles have clear and predictable behavior for every combination of $s$, $\dot{x}$, and $\dot{s}$, depending on their control design.

Generally, the problem of finding optimal parameter values to describe the car following behavior of an ACC system is posed as an optimization problem where the best-fit parameters set is the one that minimizes some error metric that measures the agreement between the calibrated model simulation and the experimental data. The question of how best to calibrate accurate microscopic models for ACC vehicles has been the focus of some recent work. Specifically, Milan\'es and Shladover~\cite{milanes2014modeling} used the mean absolute speed error to calibrate an optimal velocity type model as well as the IDM. Similarly, Gunter, et al~\cite{gunter2019model} used the root mean square error in speed as the error measure to calibrate \edit{the OVRV} car following model, and used this model to assess the string stability of the ACC dynamics. 

\edit{In this work, close the gap on the choice of target performance metric to calibrate ACC models based on trajectory data. Specifically, we investigate how best to calibrate car following models for ACC vehicle behavior by using a multi-objective approach that attempts to minimize both metrics concomitantly. While these results are in line with other recent findings on model calibration (e.g., \cite{punzo2016speed}), the results are applied specifically to ACC vehicle dynamics, which are characterized by their predictable response to lead vehicle driving behavior. The results are confirmed by calibrating different combination of models and vehicles.} 

This work builds on prior efforts by Gunter, et. al~\cite{gunter2019model,gunter2019are} and identifies how best to calibrate such car following models for commercially available ACC vehicles. The best-fit models are also presented and provide an additional set of calibrated car following models for commercially available ACC vehicles. \edit{The resulting calibrated model parameter values are specific to the vehicles tested and provide a sampling of possible ACC car following behaviors.}

The remainder of this article is outlined as follows. In Section~\ref{sec:background}, we provide a literature review on calibration metrics of car-following model. The multi-objective calibration approach used is outlined in Section~\ref{sec:multi_obj_calibration} and the three car following models considered are reviewed in Section~\ref{sec:cfmodels}. The experimental data used for calibration and testing is presented in Section~\ref{sec:data} and the resulting calibrated parameter values are presented in Section~\ref{sec:results}. The results of the validation of the calibrated models against a different dataset are presented in Section~\ref{sec:validation}. We conclude that the best fitting car following models for commercially available ACC vehicles are those calibrated to match the spacing error in Section~\ref{sec:conclusions}

\section{Background}\label{sec:background}
The calibration process is an essential step in modeling car following behavior of ACC vehicles (e.g., \cite{milanes2014modeling,gunter2019are}) as the specific control-law within the vehicle controller is unknown. Therefore, it is necessary to construct a model that is able to reproduce the traffic-level behavior of the ACC vehicle based on the observed inputs and outputs. Note that this method is not attempting to reverse engineer the actual controller being implemented on the ACC vehicle. Instead, the purpose of the model calibration is to find model parameter values that are able to reproduce the observed trajectories as accurately as possible.

However, choosing the best metric to quantify how well a particular model is able to reproduce the observed data is not straightforward. Different metrics have been proposed and used in the literature but it is not clear what are the most suitable metric for each situation. For example, \cite{milanes2014modeling} defines the average absolute speed error as the key metric for calibration parameters. The models calibrated by \cite{gunter2019are} consider the speed root mean square error. Both of them, therefore, consider speed related variables.

Various studies have defined as objective metric the spacing or position error. In particular, \cite{kesting2008calibrating} observed advantages when targeting spacing error as oppose to speed error. They observed an asymmetrical behavior with targeting spacing error in general leading to small speed errors whereas targeting speed error as the objective can lead to large position errors. More recently, an analytical study has shown a relationship between residuals of variables linked by derivatives such as speed with position and flow with cumulative flow \cite{punzo2016speed}. Specifically, they show that the sum of squared errors of the variable of the highest order (cumulative flow or position) can be decomposed into different terms and one of them is the sum of squared errors with respect with the variable of lower order (flow or speed). In other words, in order to reduce spacing error, it is also necessary to reduce speed error. This suggests that calibrating against the variable of highest order would be preferable since that leads to reduction in the errors with respect to the variable of lower order.

\edit{While there is substantial evidence of the advantage of prioritizing spacing error over speed error, many recent works, especially related to ACC vehicles, use speed error as a calibration metric. Specifically, to the best of our knowledge, no previous study has attempted to calibrate ACC vehicle car following models over the two metrics employing a multi-objective approach. One benefit of a multi-objective approach is being able to construct a Pareto set or trade-off curve, which provides insight into the relationship between the differing objectives. Assessing the trade-off curve allows us to evaluate which metric is more suitable and what is the impact (predicted error) associated with selecting a different objective for calibration.}

\edit{The Pareto set also offers a consistent method to compare multiple models across different metrics. In this work we consider the difference between calibrating ACC car following models to minimize spacing and speeds errors with respect to the experimental data, but the methodology can be utilized for other metrics as well. One benefit of this approach is that it implicitly avoids arbitrary weights for different objectives.}

\section{Multi-objective calibration approach}\label{sec:multi_obj_calibration}
\edit{
In this work we use a multi-objective optimization approach to select the model parameter values that best reproduce the experimental data in simulation. Formally, the calibration of a particular microscopic model from experimental data can be written as an optimization problem as follows: }
\begin{equation}
\begin{split}
    &min._{p_1,...,p_J} \quad J_m (\mathbf{v},\mathbf{s},\widehat{\mathbf{v}},\widehat{\mathbf{s}}) \quad m=1,...,M. \quad s.t. \\
    &\widehat{\mathbf{v}},\widehat{\mathbf{s}}  =  f(p_1,...,p_J) \\  
     &p_j^{min} \leq p_j \leq p_j^{max}  \quad j=1,...,J,
\end{split}
\label{eq:problem}
\end{equation} 
\edit{
where $M$ and $J$ are the number of objectives and model parameters respectively, $v=\dot{x}$ denotes speed, $p_1,...,p_J$ are different model parameters which should each be between $p_j^{min}$ and $p_j^{max}$ and $f(p_1,..., p_j)$ is the car following model for arguments $p_1,..., p_j$. Specifically, the function $f$ returns a set of time-series time series points of simulated speeds ($v=\dot{x}(t)$) and spacing ($s(t))$. Within the optimization problem, all values are sampled at discrete times (i.e., $v(i\Delta t)$ $s(i\Delta t)$ for integer $i$) where $\Delta t$ is the discretized time step. Finally, $J_m(.)$ are performance metrics based on the simulated outputs and measured values.}

\edit{
Unlike single-objective optimization problem in which the output is a single point, the solution to the problem \eqref{eq:problem} is a Pareto set or Pareto frontier. In the case of two objectives, the Pareto set is a line, while a pareto surface is obtained for higher-dimensional objectives. Algorithms suitable for this type of problem commonly return a set of points (solutions) along the Pareto set.}

\edit{To solve this problem, we apply the \textit{differential evolution} algorithm~\cite{storn1997differential} with the multi-objective extension \cite{robivc2005differential}. The algorithm starts by creating a population $N$ candidate solutions randomly (i.e., with values between $p_j^{min}$ and $p_j^{min}$ $\forall j$. Then at every iteration, three steps occurs:
\begin{enumerate}
    \item For every candidate solution $n$ out of $N$, create a new candidate solution based on the parent $n$ and other $p$ randomly assigned candidates ($p=2$ in our case and labeled as $i1$ and $i2$).
    \item Evaluate the new candidate by applying the function $f$ and obtaining $J_1,...,J_m$ for all new candidates.
    \item Select $N$ candidates out of $2N$ candidates (the initial $N$ added of $N$ created at step 1) through a selection process. 
\end{enumerate}
Here, the step 1 is obtained via the \textit{DE/rand1} strategy \cite{mendes2005dynde} which creates the new candidate $j$ as $\mathbf{p}_{j}:= \mathbf{p}_{j}+F(\mathbf{p}_{i1}-\mathbf{p}_{i2})$ where $i1$ and $i2$ are two randomly assigned candidates and the scalar $F$ on the interval (0,2] is a constant factor to control the amplification of the change. The step $3$ is commonly straightforward on single-objective (the candidate is kept if it is better than its parent), but in the multi-objective approach the comparison is not as evident since the new candidate solution can be better in some but not all objective compared to its parent. Therefore, additional steps are necessary.}

\edit{
Instead of a comparison with its parent, the selection is carried out by analyzing the $2N$ candidates concurrently attempting to keep "non-dominated" candidates. A candidates $i$ yielding $J_1,...,J_m$ is said to be non-dominated if there is no other candidate $j \neq i$ that yields lower value in all $m$ objectives. A non-dominated candidate is assigned rank 1. Candidates that are not rank 1, but are not dominated by other candidates except those labeled as rank 1 are assigned rank 2. Similar for higher ranks until all candidates are assigned a rank. The selection occurs by taking the best $N$ candidates with respect to their rank. In case of the high ranking considered exceed $N$ candidates, the candidates within that rank is selected through the Crowding Distance sorting \cite{deb2002fast}, which attempts to keep ``diverse'' candidates. For specific details we refer to  \cite{robivc2005differential} and \cite{desouza2019}. The final goal is to obtain $N$ rank 1 candidate solutions.}

\edit{
The algorithm runs iteratively until a convergence criteria is met or sufficient iterations are completed to satisfy convergence (i.e., run beyond convergence).}

\section{Car following models}
\label{sec:cfmodels}

We evaluate three car-following models: \edit{the OVRV model, which is a form of an optimal velocity model~\cite{milanes2014modeling}}, the Intelligent Driver model \cite{treiber2000congested}, and the Gipps' model \cite{gipps1981behavioural}. We review the model assumptions as well as the assumptions undertaken to run the numerical scenarios in the same step as the data available.

Time is denoted as $t$ and time is discretized into fixed steps $\Delta t$ such that $t=i\Delta t$ with $i$ the discrete index. \edit{State variables} position ($x$) and speed ($v$) are related by the discretization scheme for position update:
\edit{
\begin{equation}
    x(t+\Delta t) = x(t) + \frac{1}{2} (v(t) + v(t+\Delta t)) \Delta t,
\end{equation}}
\edit{where }the implicit assumption is that the vehicle reaches speed $v(t+\Delta t)$ at the end of the time-step. The speed $v(t)$ is defined by the specific car-following model. Also, we assume time-steps at the same frequency as the data available for all models. To compute position, spacing and speeds at continuous time $t-\tau$, we use an interpolation method \cite{kesting2008reaction} with variable $z$ at time $t-\tau$ computed as:
\begin{equation}
z(t-\tau) = \beta z(t-(n+1)\Delta t) + (1-\beta)z(t-n\Delta t)
\label{eq:interpolation}
\end{equation}
where $n$ constitutes the integer part of $\tau/\Delta t$ and $\beta = \tau/\Delta t - n$ which is necessary for the models with arbitrary delay, \edit{such as the Gipps model,} that are not necessarily multiples of the time-step assumed. The speed and position at time $t$ for the lead vehicle are denoted as $v_\mathrm{L}(t)$, $x_\mathrm{L}(t)$, respectively. We define the bumper to bumper spacing $s(t)$
\begin{equation}
    s(t) = x_\mathrm{L}(t)-x(t) - L,
\end{equation}
where $L$ is the vehicle length. Commonly when a vehicle is equipped with sensors, it is able to measure spacing $s(t)$ as oppose the front rear to front rear spacing ($x_\mathrm{L}(t)-x(t)$). This will be the case for our test data. In this case, for converting the spacing data into absolute position, it is necessary to consider the vehicle length. We assumed $L$ to be $4.8$m for all the numerical experiments in which this metric was necessary.

The three car-following models considered are presented in the following sub-sections.
\subsection{Optimal Velocity Relative Velocity model}
\edit{The OVRV is based on the commonly assumed \textit{constant time gap} (CTG) control law for ACC vehicles~\cite{rajamani1998design}, and is often used to describe ACC vehicle dynamics~\cite{gunter2019are,milanes2014modeling}:}
\begin{equation}
    \begin{split}
        \dot{v}(t) = k_1[s(t)-\eta-t_hv(t)] + k_2(v(t)-v_\mathrm{L}(t)),
    \end{split}
\end{equation}
where $\eta$ is the jam space gap, $t_h$ is the desired time gap with respect to the leader. Parameters $k_1$ and $k_2$ are, respectively, the spacing and speed gains. \edit{Note that this is an extension of the OVM~\cite{bando1995dynamical} with an optimal velocity law $v = (s-\eta)/t_h$.} It is common to assume a delay between a change in speed and spacing and the response of the following vehicle. To account for delays, we assume spacing and leading speed measurements are subject to time delay $\tau$ resulting in the following speed dynamics \cite{gunter2019are}:
\begin{equation}
    \begin{split}
        \dot{v}(t) = k_1[s(t-\tau)-\eta-t_h v(t)] + k_2(v(t)-v_\mathrm{L}(t-\tau)),
    \end{split}
\end{equation}
For more details on this model we refer to \cite{bando1995dynamical} and \cite{gunter2019are}.

\subsection{Gipps}
Gipps car-following model \cite{gipps1981behavioural} is derived \edit{on the concept of} safe-following distance. The basic assumption is that the driver will always leave enough gap to its leader vehicle so as to avoid a collision on the worst case scenario - the leader breaks at maximum deceleration rate $\hat{B}_L$. It is also assumed that the driver always reacts after a reaction time $\tau$ and also consider an addition time safe margin $\theta$. Moreover, the driver is able to brake at deceleration rate $B$. When not constrained by the leading vehicle, the drivers follow an acceleration-speed relationship normalized by a maximum acceleration, $A$ (with constants defined for the metric system):
\begin{subequations}
\begin{equation}
v(t)= \min \{v^{\textrm{u}}(t), v^{\textrm{c}}(t) \}
\end{equation}
\begin{equation}
v^{\textrm{u}}(t) = v(t-\tau ) + 2.5A\tau \bigg(1-\frac{v(t-\tau)}{V}\bigg)\sqrt{0.025 + \frac{v(t-\tau)}{V}},
\end{equation}
\begin{equation}
\begin{split}
v^{\textrm{c}}(t) = & \max \bigg\{\bigg(B^2(\frac{\tau}{2}+\theta)^2 + B \{2(x_\textrm{L}(t-\tau ) \\& - x(t-\tau) -\eta - L ) -\tau v(t-\tau) + \frac{v_\textrm{L}(t-\tau )^2}{\widehat{B}_\textrm{L}} \} \bigg)^{1/2} \\&-B (\frac{\tau}{2}+\theta), \\ &v(t-\Delta t) - B\Delta t \bigg\}.
\end{split}
\label{eq:gippsconstrained}
\end{equation}
\label{eq:gipps}
\end{subequations}
where $v^{\textrm{c}}(t)$ signifies the speed constrained by the leader vehicle, $v^{\textrm{u}}(t)$ denotes the speed when movement is not constrained by the leading vehicle, $V$ is the free-flow speed; $S_L$ characterizes the effective length of the leader, $\tau $ is the reaction time, and $\theta$ is the safety margin. Gipps model assumes a single parameter for the jam spacing which is the sum of the physical length and a safe margin. Here we keep consistency with the other models used and use spacing gap $\eta$ added of the physical length, $L$ instead of the jam spacing in the original model. 

\edit{The Gipps model assumes a }time-step equal to the reaction time (i.e., $\Delta t=\tau$) and therefore all variables depend in the previous time step. This is the reason we added the the second input to the max operator in Equation~\ref{eq:gippsconstrained} which ensures the subject vehicle will not decelerate at a rate higher than $B$.

Recall that whenever $\frac{\tau}{\Delta t}$ is not integer, the interpolation as Equation \ref{eq:interpolation} is applied.

\subsection{Intelligent Driver model}

Finally, we consider the Intelligent Driver model \cite{treiber2000congested}. Unlike the Gipps model, the IDM model does not have two distinct regimes for speed update. Instead, the acceleration and deceleration behavior are considered concurrently. The state update is as follows:
\begin{equation}
\begin{split}
    \dot{v}(t) &= A\bigg[1- \bigg(\frac{v(t)}{V} \bigg)^\delta - \bigg( \frac{s^\star(t)}{s(t)} \bigg)^2 \bigg], \\
    s^\star(t) &= \eta + v(t)t_h + \frac{v(t)(v(t)-v_\textrm{L}(t)\edit{)}}{2\sqrt{AB}}
\end{split}
\end{equation}
where $A$ \edit{is the maximum acceleration rate  and $B$ if the comfortable deceleration rate}, $\eta$ is the jam spacing, $\delta$ is a free acceleration exponent, $t_h$ the desired time-gap and $V$ the desired speed. The term $s^\star$ represents an effective gap given the prevailing conditions. The steady state desired gap can be obtained assuming $v(t)=v_\textrm{L}(t)=v$ and it becomes $s = \eta + vt_h$.

\section{Calibration data}\label{sec:data}
The experimental data used in this study was collected during a series of two-vehicle car following experiments conducted with seven different commercially available ACC vehicles. The data was collected by Gunter, et al.~\cite{gunter2019are} and contains over 1,200 miles of car following data. These vehicles (A-G) range from a small hybrid hatchback vehicle to a large internal combustion engine SUV with the same vehicle designation (A-G) used for each vehicle as in the original dataset published by Gunter, et al.~\cite{gunter2019are}.

Each experiment involves two vehicles: a lead vehicle \edit{and a following vehicle. The lead vehicle is unchanged for all tests and is followed by one of the seven vehicles tested} 
that drives behind the lead vehicle with ACC engaged. A total of four different tests were conducted for each vehicle. The tests were designed to capture a range of ACC following behavior and include both constant speed portions at different speeds as well as oscillatory behavior and rapid braking events. A full description of all tests conducted is presented below:

\begin{itemize}
    \item \textbf{Low-speed speed step:} The lead increases speed from 15.6 m/s (35 mph) mph to 24.6 m/s (55 mph) in 2.23 m/s (5 mph) increments, holding each speed for 60 seconds in a step-wise manner before reducing speeds back to 15.6m/s (35 mph) in 2.23 m/s (5 mph) increments, holding each speed for 60 seconds again.
    
    \item \textbf{High-speed speed step:} The lead increases speed from 26.8 m/s (60 mph) to 31.2 m/s (70 mph) in 2.23 m/s (5 mph) increments, holding each speed for 60 seconds in a step-wise manner before reducing speeds back to 26.8 m/s (60 mph) in 2.23 m/s (5 mph) increments, holding each speed for 60 seconds again.
    
    \item \textbf{Oscillatory:} The lead vehicle fluctuates speed between 24.6 m/s(55 mph) and 21.9 m/s (49 mph) for the first half of the test, holding each speed for 30 seconds, and fluctuates between 24.6 m/s (55 mph) and 20.1 m/s (45 mph), again holding each speed for 30 seconds for the second half of the test.
    
    \item \textbf{Speed dips:} The lead vehicle drives at 24.6 m/s (55 mph), holding this speed for at least 45 seconds before \edit{quickly braking for several seconds before returning to 24.6 m/s (55 mph) for at least 45 seconds after each speed dip. The magnitude of the braking is varied for each dip (2.24 m/s (5 mph), 4.47 m/s (10 mph), and 6.70 m/s (15 mph) dips are tested).}
\end{itemize}

In every vehicle, it is possible to adjust the desired following gap to the leading vehicle between the closest (min) following setting and furthest (max) following setting. Experiments were conducted twice, one for each following setting..

\section{Calibration results}\label{sec:results}
We apply the multi-objective differential evolution \cite{storn1997differential,deb2002fast} with same setting as \cite{desouza2019} to calibrate the three models introduced in Section \ref{sec:cfmodels}. We consider the spacing error, $s_e$, and the speed error, $v_e$ as the objectives:
\begin{equation}
\begin{split}
    s_e = \sqrt{\sum_{i=1}^I \frac{(\hat{s}(i\Delta t)-s(i\Delta t))^2}{I} }  \\
    v_e = \sqrt{\sum_{i=1}^I \frac{(\hat{v}(i\Delta t)-v(i\Delta t))^2}{I} },  \\
\end{split}
\end{equation}
where $I$ is the number of samples for each test. Measurements are available at time-step of $0.1$ s and we run the numerical experiments with the same time-step (i.e., $\Delta t=0.1$ s). Table~\ref{tab:modelparams} shows the model parameters for all three models, and their respective acceptable range in values.

\begin{table}[]
    \centering
    \begin{tabular}{c c c c c}
     \hline
        Parameter & Model & Description              & Unit  & Range  \\ \hline
        $\eta$      & OVRV & Jam spacing              & m     & \edit{0}-17 \\ 
        $t_h$     & OVRV  & Desired time-gap         & s     & 0-2.5  \\ 
        $\theta$    & OVRV  & Sensor delay             & s     & 0-2.5  \\  
        $k_1$       & OVRV  & spacing gain             & 1/s & 0-0.3  \\  
        $k_2$       & OVRV   & speed gain               & 1/s\textsuperscript{2} & 0-0.6  \\  
        $A$        & Gipps & Max. acceleration  & m/s\textsuperscript{2} & 0-5  \\   
        $B$        & Gipps & Max. deceleration & m/s\textsuperscript{2} & 2-9  \\   
        $B_{avg}$     & Gipps & Max. deceleration & m/s\textsuperscript{2} & 2-9   \\  
        $\eta$        & Gipps & Jam spacing   & m     & 0-\edit{17} \\   
        $\tau$      & Gipps & reaction time      & s     & 0.1-2 \\   
        $\theta$     & Gipps & safe margin        & s     & 0-2   \\  
        $V$         & Gipps & Desired speed      & m/s   & 25-40 \\  
        $V$         & IDM & Desired Speed         & m/s   & 25-40  \\  
        $\delta$     & IDM & Free-flow exponent     & -     & 0.2-20 \\  
        $t_h$         & IDM & Desired time gap      & s     & 0-2.5    \\  
        $\eta$         & IDM & jam distance          & m     & 0-\edit{17}    \\  
        $A$         & IDM & Maximum acceleration  & m/s\textsuperscript{2} & 0-5    \\  
        $B$         & IDM & \edit{Comfortable} deceleration & m/s\textsuperscript{2} & 2-9    \\ \hline
    \end{tabular}
    \caption{Model parameters and respective ranges.}
    \label{tab:modelparams}
\end{table}

For each vehicle and spacing setting, \edit{we used the the first half of all driving cycles (low and high speed step, oscillatory, and speed dips) as calibration data. To avoid any bias related to the initial state response, we disregard the first 10\% of the data. The second half of each driving cycle is is kept for validation assessment.} The multi-objective approach \cite{desouza2019} and described in Section \ref{sec:multi_obj_calibration} is applied to each combination of vehicle and ACC setting (e.g., Vehicle A, min setting, Vehicle A, max setting, etc.) for the different car-following models. For all calibration runs, a population size $N=100$ (that is, the output contains 100 different solutions yielding different $s_e$ and $v_e$). \edit{In all cases, the Pareto set does not change significantly after few hundred iterations. Based on that, all cases are run for 1000 iterations, which is sufficient for convergence in all cases.}

\begin{figure}[!h]
    \centering 
\begin{minipage}{0.45\textwidth}
  \includegraphics[width=\linewidth]{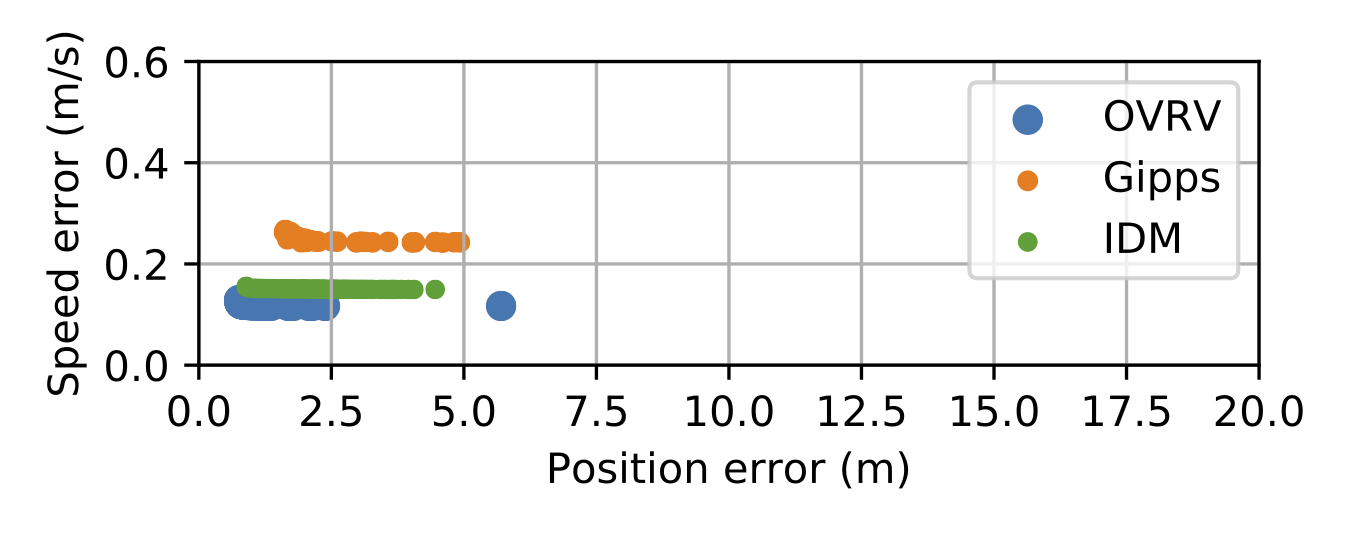}
  \subcaption{Pareto frontier for vehicle A under the minimum following setting.}
  \label{fig:8}
\end{minipage} 
\begin{minipage}{0.45\textwidth}
  \includegraphics[width=\linewidth]{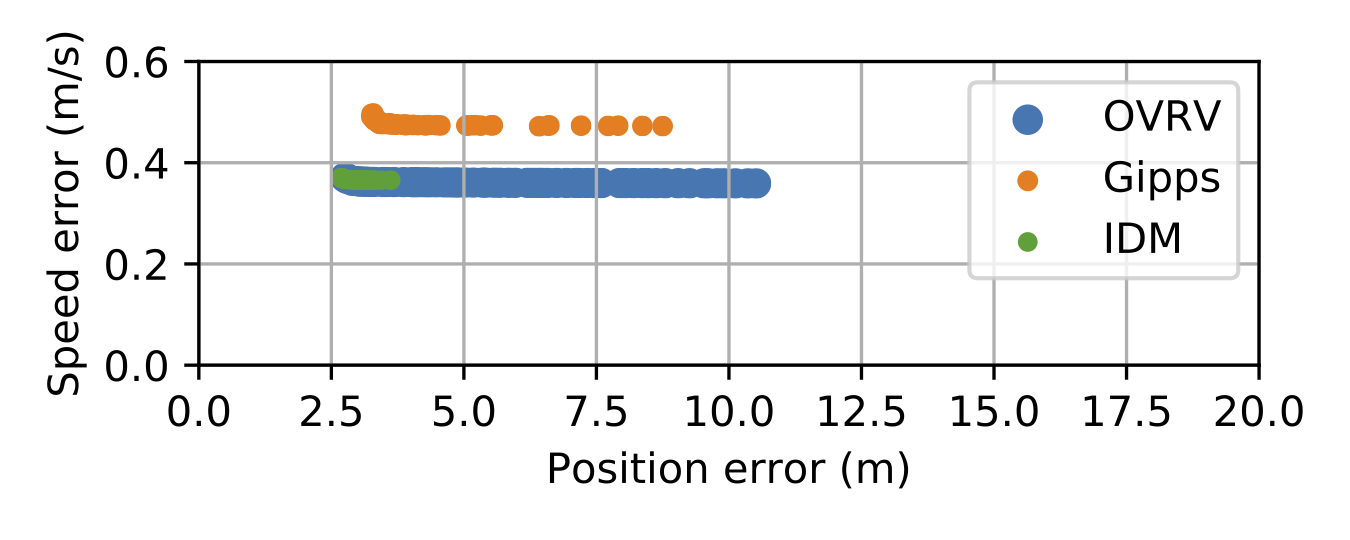}
  \subcaption{Pareto frontier for vehicle B under the minimum following setting}
  \label{fig:9}
\end{minipage} 
\begin{minipage}{0.45\textwidth}
  \includegraphics[width=\linewidth]{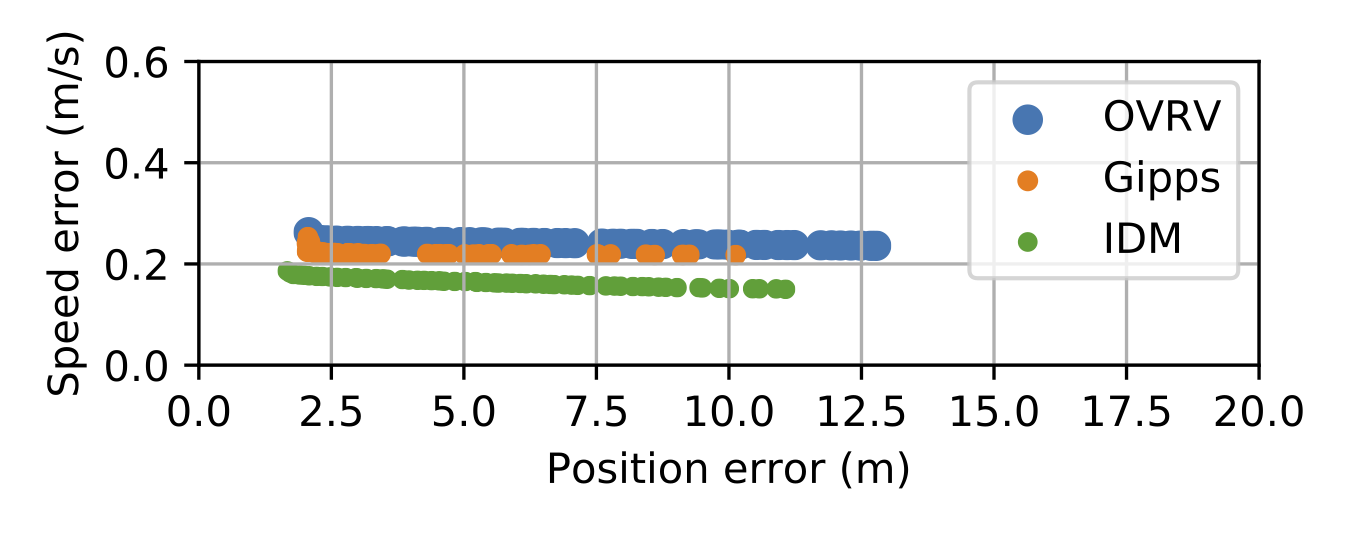}
  \subcaption{Pareto frontier for vehicle C under the minimum following setting.}
  \label{fig:10}
\end{minipage}
\begin{minipage}{0.45\textwidth}
  \includegraphics[width=\linewidth]{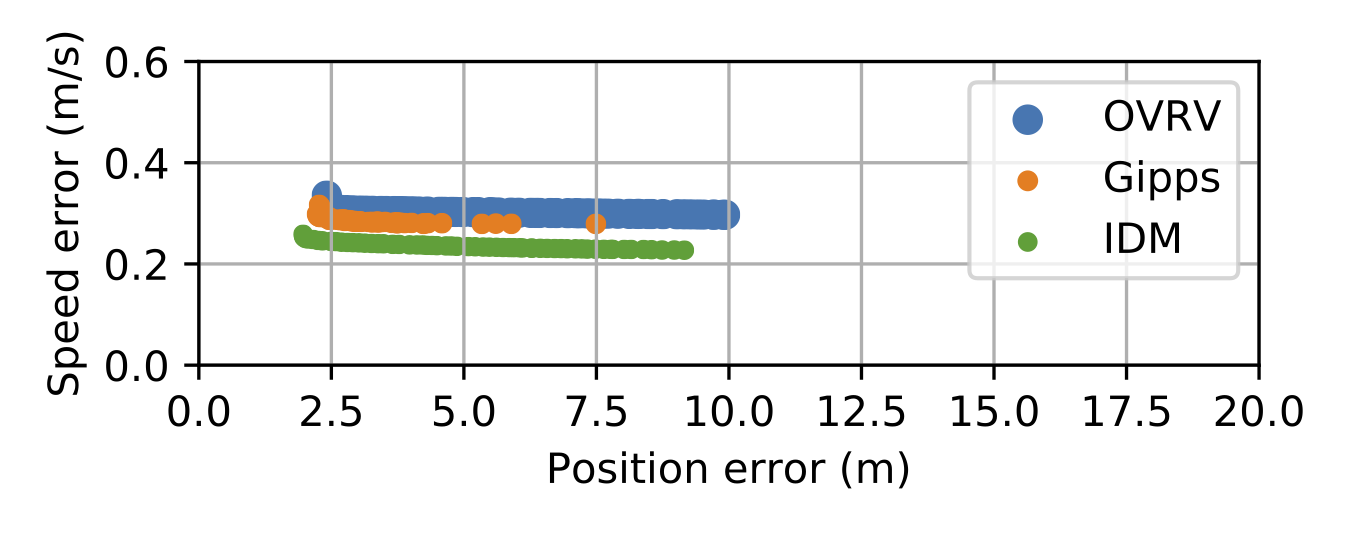}
  \subcaption{Pareto frontier for vehicle D under the minimum following setting.}
  \label{fig:11}
\end{minipage}\hfil 
\centering
\begin{minipage}{0.45\textwidth}
  \includegraphics[width=\linewidth]{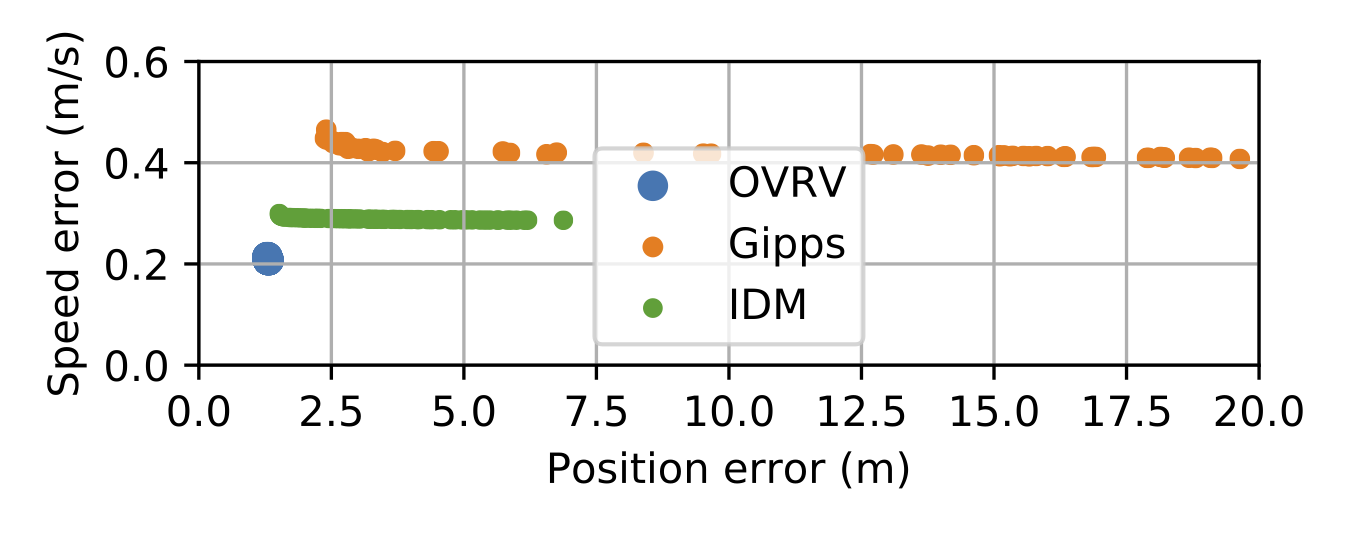}
  \subcaption{Pareto frontier for vehicle E under the minimum following setting.}
  \label{fig:12}
\end{minipage} 
\begin{minipage}{0.45\textwidth}
  \includegraphics[width=\linewidth]{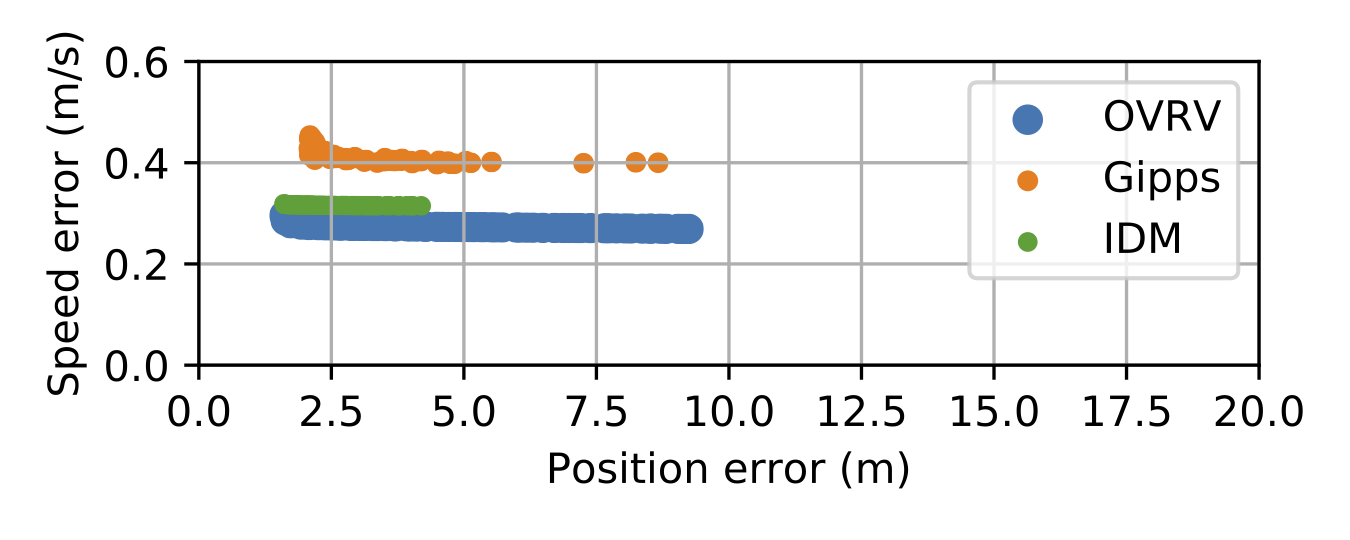}
  \subcaption{Pareto frontier for vehicle F under the minimum following setting.}
  \label{fig:13}
\end{minipage} 
\begin{minipage}{0.45\textwidth}
  \includegraphics[width=\linewidth]{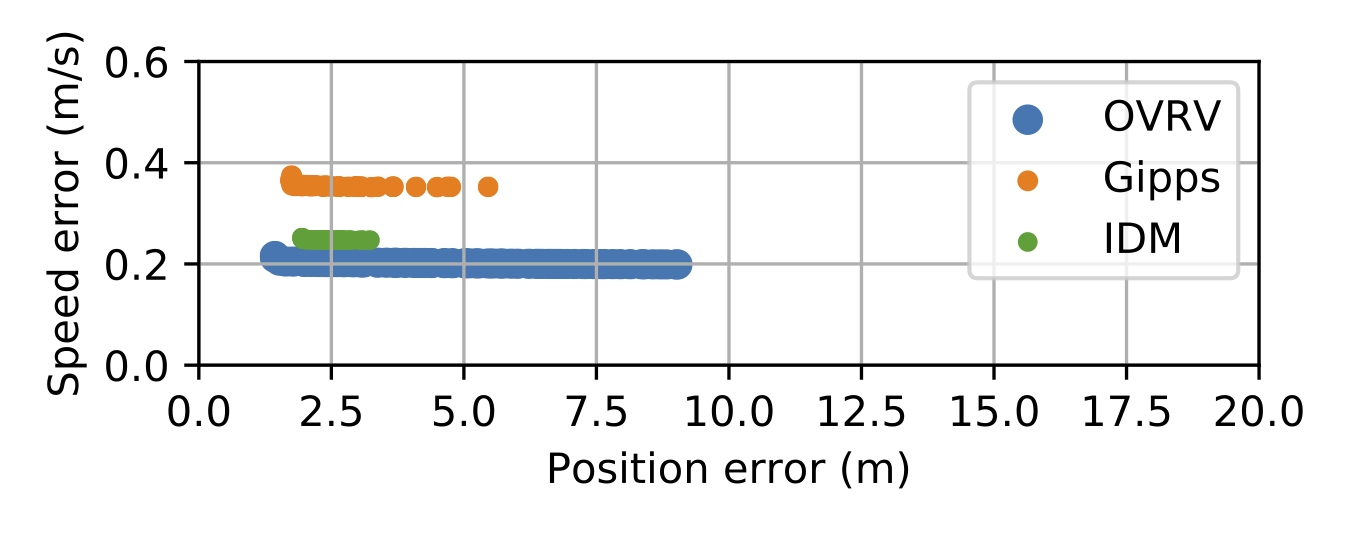}
  \subcaption{Pareto frontier for vehicle G under the minimum following setting.}
  \label{fig:14}
\end{minipage}
\caption{Pareto frontier for all vehicles using data for the minimum following setting for the OVRV (blue), Gipps (orange), and IDM (green). The X-axis and Y-axis represents RMSE of spacing and speed error, respectively. \edit{Note that some points on the Pareto frontier overlap, but each model is calibrated with the same number of trials.}}
\label{fig:paretomin}
\end{figure}

Figures \ref{fig:8} to \ref{fig:14} show the Pareto frontier for vehicles A trough G for the minimum setting. The Pareto frontier for calibrating models based on data collected at the maximum setting is not shown since it follows the same trends observed in Figures~\ref{fig:8} to~\ref{fig:14}. In all graphs, the x-axis represents the spacing RMSE and y-axis the speed RMSE. Each dot represents a specific solution in the Pareto frontier. The blue dots are associated to the OVRV, the green dots IDM, and the orange dots to the Gipps' model.    

A model can be regarded as superior if it outperforms all the other models (for minimization, if it consistently produces lower error and thus lies below the other curves) across all objectives in the domain. In general, \edit{OVRV and IDM yield} comparable results with, perhaps, a slight advantage of the former. OVRV outperforms IDM for vehicles A, E, F, and G on both ACC vehicle following settings (maximum setting not shown in Figures~\ref{fig:8} to~\ref{fig:14}) whereas IDM outperformed for vehicles C and D. \edit{The different performance of each model when modeling different ACC vehicles is a result of the different vehicle-level dynamics, and which model they more closely resemble. Importantly, the OVRV is based on the constant time gap control law, which is used for many ACC vehicles, and explains the relatively good performance of the OVRV.} Gipps' model is generally outperformed by the other car-following models, with few exceptions. \edit{It is worth mentioning that relative performance may be sensitive to the data subset used for calibration, especially at different regimes as pointed out in \cite{sharma2019more}. Although we do not perform the same analysis, we test calibration using only the step-high and step-low driving cycles and we also found a comparable performance between OVRV and IDM model.}

We would like to point out two aspects of the Gipps model. First, it is important to note that the model was not designed for short time-steps such as the ones we are considering here. We adapted the equation set accordingly to our numerical experiments, but we want to \edit{clarify that the model was not designed for this type of experiments, but rather is being applied here for demonstration purposes}. Second, Gipps' car-following model has two different regimes (congested and uncongested), which leads to a speed discontinuity when switching between regimes. However, the observed ACC data has a few discontinuities, which are induced by the lead vehicle driving behavior. This is the most likely reason why Gipps model under-performs compared to the other models evaluated. Also, it is important to note that these relative performances and aspects do not necessarily transfer to human-driven vehicles.

Moreover, it is interesting to note that most of the trade-off curves have a small amplitude on speed error and larger amplitude on the spacing error, especially for OVRV and IDM models. This suggests the target should be minimizing spacing error instead of speed errors as commonly applied, e.g.,~\cite{milanes2014modeling,gunter2019are} and support previous analytical studies ~\cite{punzo2016speed} that spacing should be preferred over speed error.



\begin{figure}[ht]
    \centering
    \begin{subfigure}{0.7\textwidth}
    \centering
    \includegraphics[width=\textwidth]{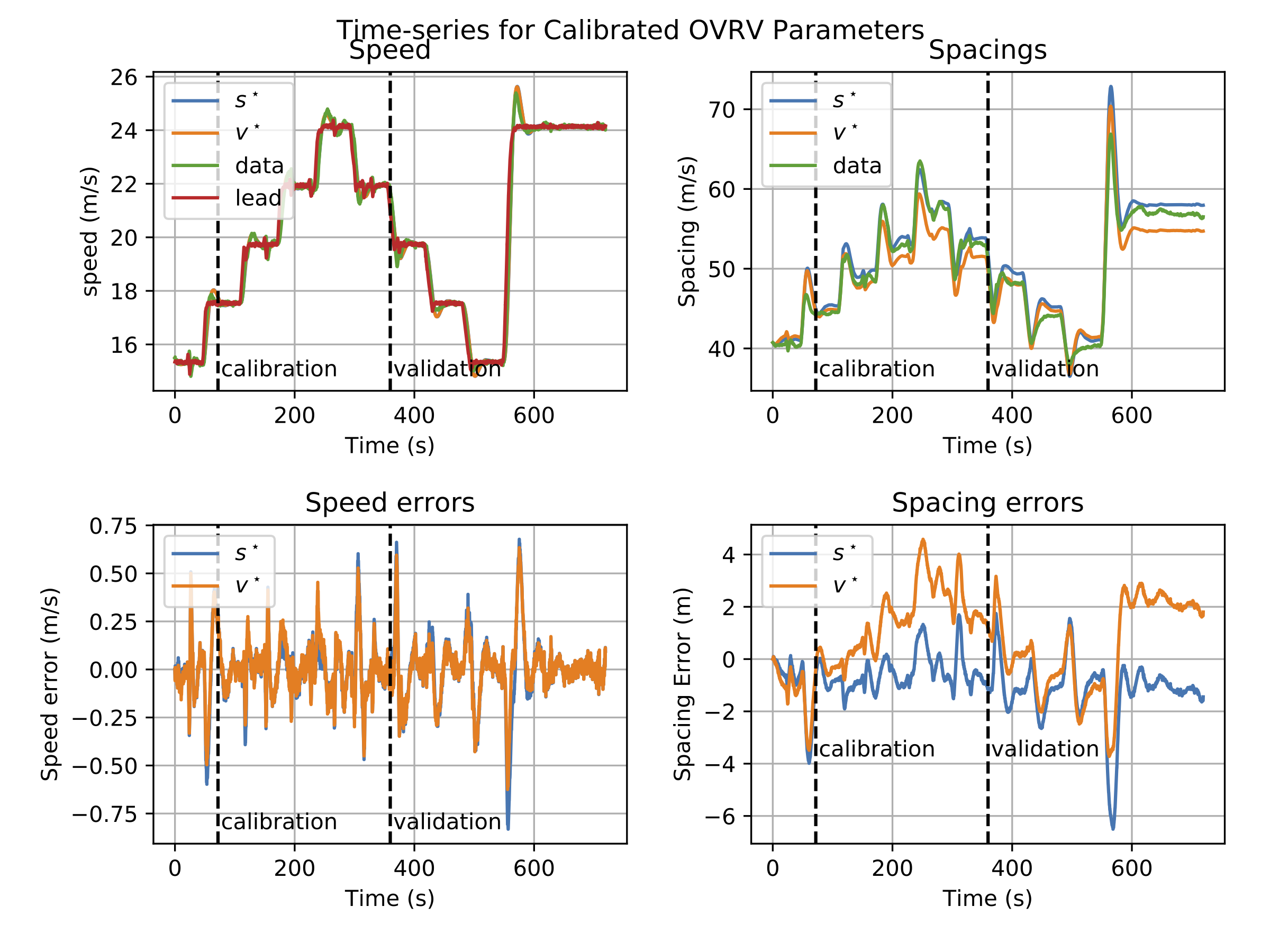}
    \caption{Validation results for OVRV.}
    \vspace{2em}
    \label{fig:timeseriesOVRV}
 \end{subfigure}\\
 \begin{subfigure}{0.7\textwidth}
    \centering
    \includegraphics[width=\textwidth]{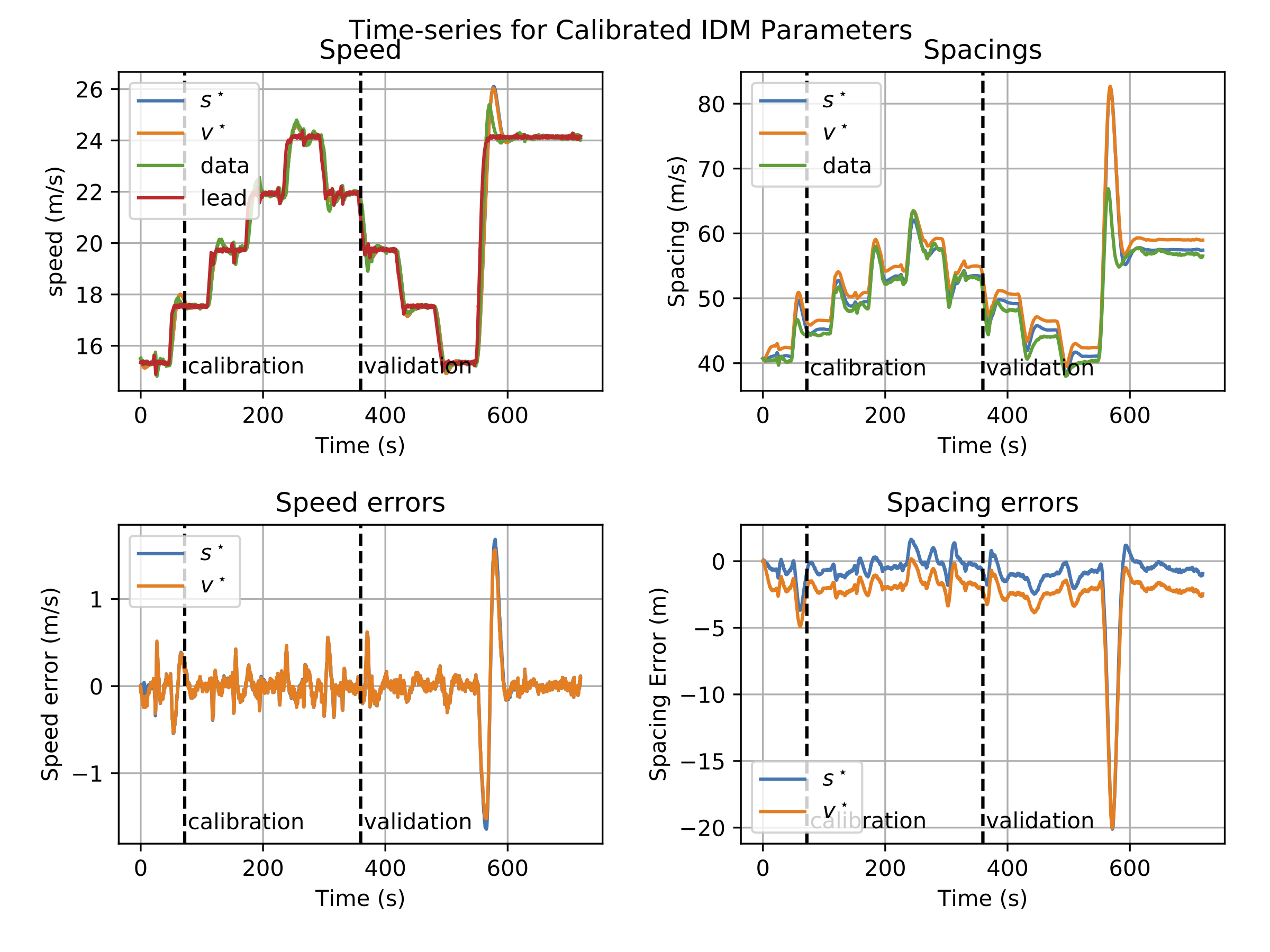}
    \caption{Validation results for IDM.}
    \vspace{2em}
    \label{fig:timeseriesidm}
 \end{subfigure}
 \caption{Clockwise from top left: time-series of (\textit{i}) speeds, (\textit{ii}) spacing, (\textit{iii}) speed errors and (\textit{iv}) spacing errors for the vehicle F for OVRV (top) and IDM (bottom). The blue lines represents the solution with smaller spacing error and the orange lines represents the solution of smaller speed error. In the speed and spacing graphs (\textit{i} and \textit{ii}) the green line represents the observed data. The lead vehicle speeds is depicted in red on graph (\textit{i}).}
\end{figure}

 
The results suggest that the solution with lower spacing error would also yield close to the minimum speed error but the opposite not holding. We can observe this behavior on Figures~\ref{fig:timeseriesOVRV} and~\ref{fig:timeseriesidm} which depict the time-series for the low-speed step test for OVRV and IDM models. In both, the top left graph depicts the lead vehicle speed in red \edit{and the actual speed} in green. The blue and orange lines depict the speeds for the calibrated models at the minimum spacing error and minimum speed error, respectively. The same color \edit{scheme} applies to the other three graphs with the top right graphs depicting spacing, bottom left \edit{depicting} speed errors ($v_e$) and the bottom right \edit{depicting} spacing error ($s_e)$. The \edit{dashed lines mark the time periods used for calibration and for validation purpose}. For both OVRV and IDM the calibrated speeds almost overlaps for the parameter set $v^\star$ and $s^\star$. However, they have distinct results regarding spacing. In particular, the OVRV appears to have a consistent bias on the spacing-speed relationship when traveling at constant speeds. This does not occur on IDM even though  it also seems to yield a bias at constant speeds, but significantly smaller.  

As the Pareto frontier's are flat, we report here only the solution with smallest spacing error. Table~\ref{tab:results_params} shows the parameters of each model. These calibrated model parameter values will prove valuable in conducting microscopic traffic simulations with ACC vehicles in the flow. Though the parameters across models have similar meaning (for example, $t_h$ for OVRV and IDM and $\tau$ for delay in control actions or reaction time, and free-flow speed, $V$) they are not expected to have the same values. For the same vehicles, the speed parameters $V$ are similar for both Gipps and IDM models. Gipps model had consistently high reaction time, $\tau$.

\begin{table*}[ht]
\begin{center}
\resizebox{\textwidth}{!}{%
\begin{tabular}{cc|ccccc|ccccccc|cccccc}
    \hline

    \multicolumn{2}{c} {} & \multicolumn{5}{c}{ OVRV} & \multicolumn{7}{c}{Gipps} & \multicolumn{6}{c}{IDM} \\
     \multicolumn{1}{c}{Vehicle} & \multicolumn{1}{c}{Setting} & $t_h$  & $\eta$ & $\tau$ & $k_1$ & $k_2$ & $\eta$ & $\tau$ & $\theta$ & $V$ & $A$ & $B$ & $B_{avg}$ & $t_h$ & $\eta$ & V & A & B & $\delta$ \\
     \multicolumn{2}{c}{Units} & $s$ & $m$ & $s$ & $\frac{1}{s}$ & $\frac{1}{s^2}$ & $m$ & $s$ & $s$ & $\frac{m}{s}$ & $\frac{m}{s^ 2}$ & $\frac{m}{s^ 2}$ & $\frac{m}{s^ 2}$ & $s$ & $m$ & $\frac{m}{s}$ &  $\frac{m}{s^ 2}$ & $\frac{m}{s^ 2}$ & $-$ \\
\hline 
A&min&1.0&9.4&0.58&0.05&0.26&0.3&1.8&0.0&37.5&0.5&2.6&2.4&1.0&8.0&43.6&0.9&9.0&13.5\\
A&max&2.1&10.5&0.59&0.02&0.15&2.0&2.0&0.6&37.0&0.4&7.9&7.0&2.2&6.3&44.1&0.6&5.2&15.5\\
B&min&0.9&10.9&0.51&0.04&0.16&0.4&1.9&0.0&35.5&0.6&3.0&2.6&1.0&8.0&42.8&0.8&9.0&16.2\\
B&max&2.0&15.6&0.59&0.03&0.09&1.9&1.9&1.3&37.2&0.3&4.1&3.3&2.3&7.9&42.4&0.7&8.9&17.5\\
C&min&0.8&15.8&0.36&0.04&0.2&1.5&1.9&0.0&37.6&0.4&2.3&2.1&1.1&7.8&43.7&0.6&8.6&17.1\\
C&max&2.0&16.8&0.24&0.02&0.12&4.4&1.3&1.9&39.7&0.2&7.0&5.1&2.3&8.0&43.6&0.6&6.5&17.1\\
D&min&0.8&14.2&0.46&0.05&0.19&0.9&1.9&0.1&36.2&0.4&3.4&2.9&1.0&8.0&43.4&0.6&9.0&14.8\\
D&max&1.9&15.9&0.49&0.02&0.12&4.6&2.0&1.0&36.7&0.4&4.4&3.7&2.2&8.0&44.7&0.7&8.7&17.3\\
E&min&1.3&4.5&0.60&0.06&0.16&0.5&1.7&0.1&35.7&0.5&2.3&2.1&1.3&4.2&42.6&0.9&9.0&18.9\\
E&max&2.0&9.2&0.60&0.06&0.11&4.0&2.0&0.5&36.3&0.5&8.8&7.6&2.0&8.0&44.8&1.3&9.0&19.6\\
F&min&0.8&12.2&0.54&0.06&0.17&0.7&1.8&0.1&36.7&0.5&4.9&4.1&1.0&8.0&44.7&0.8&9.0&19.7\\
F&max&1.9&6.1&0.60&0.04&0.13&1.0&1.9&0.4&37.6&0.4&7.9&7.1&1.9&7.1&41.9&0.8&9.0&13.6\\
G&min&0.6&16.9&0.58&0.06&0.21&0.4&2.0&0.1&37.9&0.6&3.2&2.7&1.0&8.0&44.9&0.8&9.0&19.8\\
G&max&2.1&2.6&0.59&0.04&0.12&0.5&2.0&0.2&37.8&0.4&8.0&8.3&2.0&4.1&44.0&0.9&9.0&11.1\\
    \hline
\end{tabular}}
\end{center}
\caption{Calibrated parameters for all vehicles for minimum and maximum setting for the solution of smaller spacing error.}\label{tab:results_params}
\end{table*}

\section{Platoon Validation}
\label{sec:validation}
The dataset used to calibrate models for this study published in~\cite{gunter2019are} also contains trajectories of an experiment conducted with a platoon of 7 ACC vehicles following a lead vehicle executing a pre-specified speed profile. The following vehicles all drive with ACC engaged, and the goal of this experiment is to assess how changes in the lead vehicle speed (i.e., perturbations) propagates through the platoon of ACC vehicles. The following 7 vehicles were Model A vehicles with ACC engaged following the minimum setting.

The experiment starts with the lead vehicle driving at a constant speed of 22.4 m/s (50 mph). Once all following vehicles stabilizes at the same speed, the lead vehicle decelerates to 19.6 m/s (44 mph), while all following vehicles drive under ACC. After some time, the lead vehicle accelerates to return to the original speed of 50 mph (22.4 m/s).

The calibrated car-following models for that specific model and ACC setting were used to replicate the same experiment in simulation given the lead vehicle trajectory. The three car-following models are applied for two parameters set on the two extremes of the Pareto set - the solution of minimum spacing error, $s^\star$,  and the solution of the minimum speed error, $v^\star$.

\begin{figure}
    \centering
    \begin{minipage}{0.39\textwidth}
  \includegraphics[width=\textwidth]{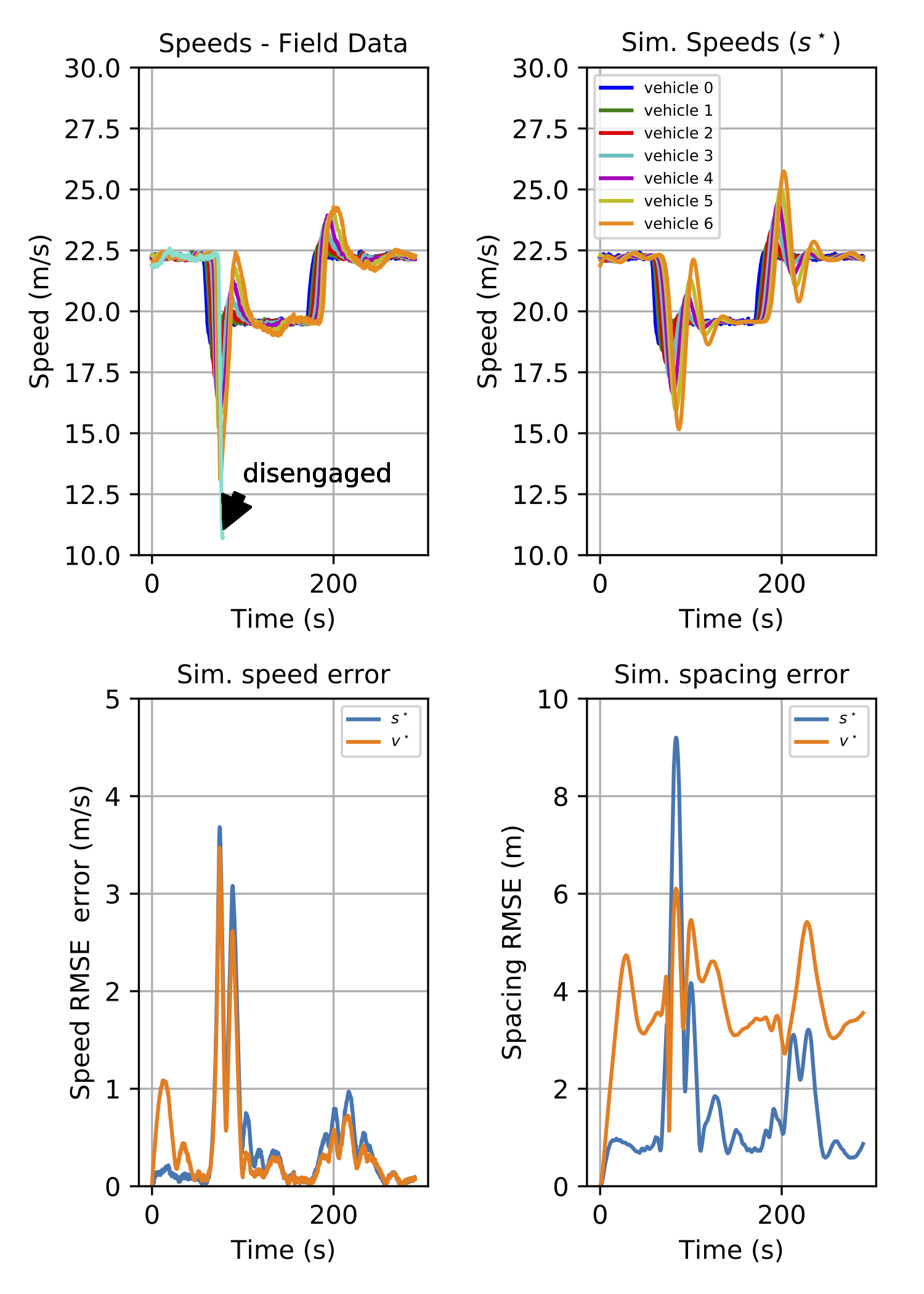}
  \subcaption{Speed and position error for the OVRV model.}
  \label{fig:platoonovm}
\end{minipage}
\qquad
\begin{minipage}{0.39\textwidth}
\includegraphics[width=\textwidth]{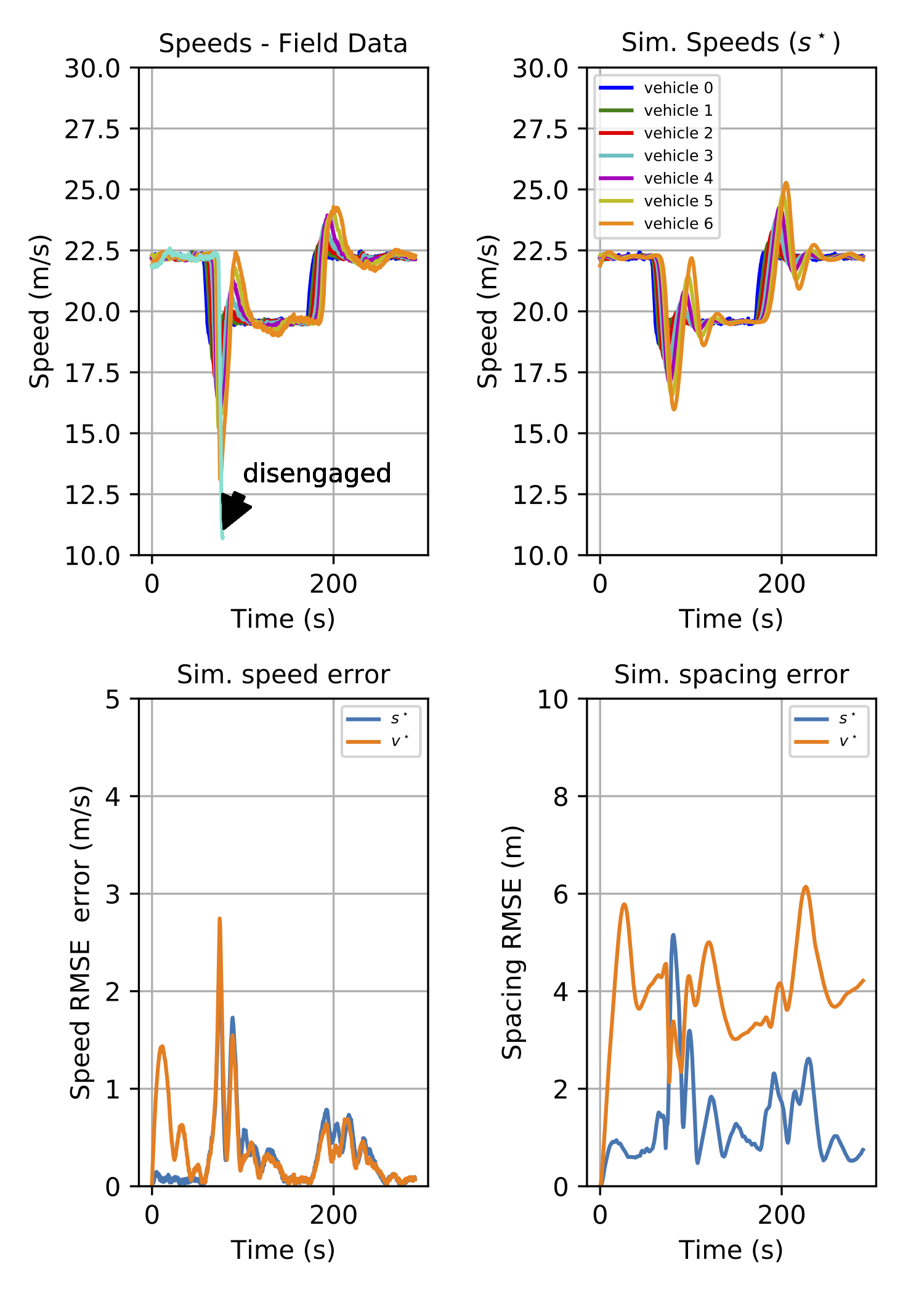}
  \subcaption{Speed and position error for the IDM model.}
  \label{fig:platoonidm}
\end{minipage}
\qquad
\begin{minipage}{0.39\textwidth}
  \includegraphics[width=\textwidth]{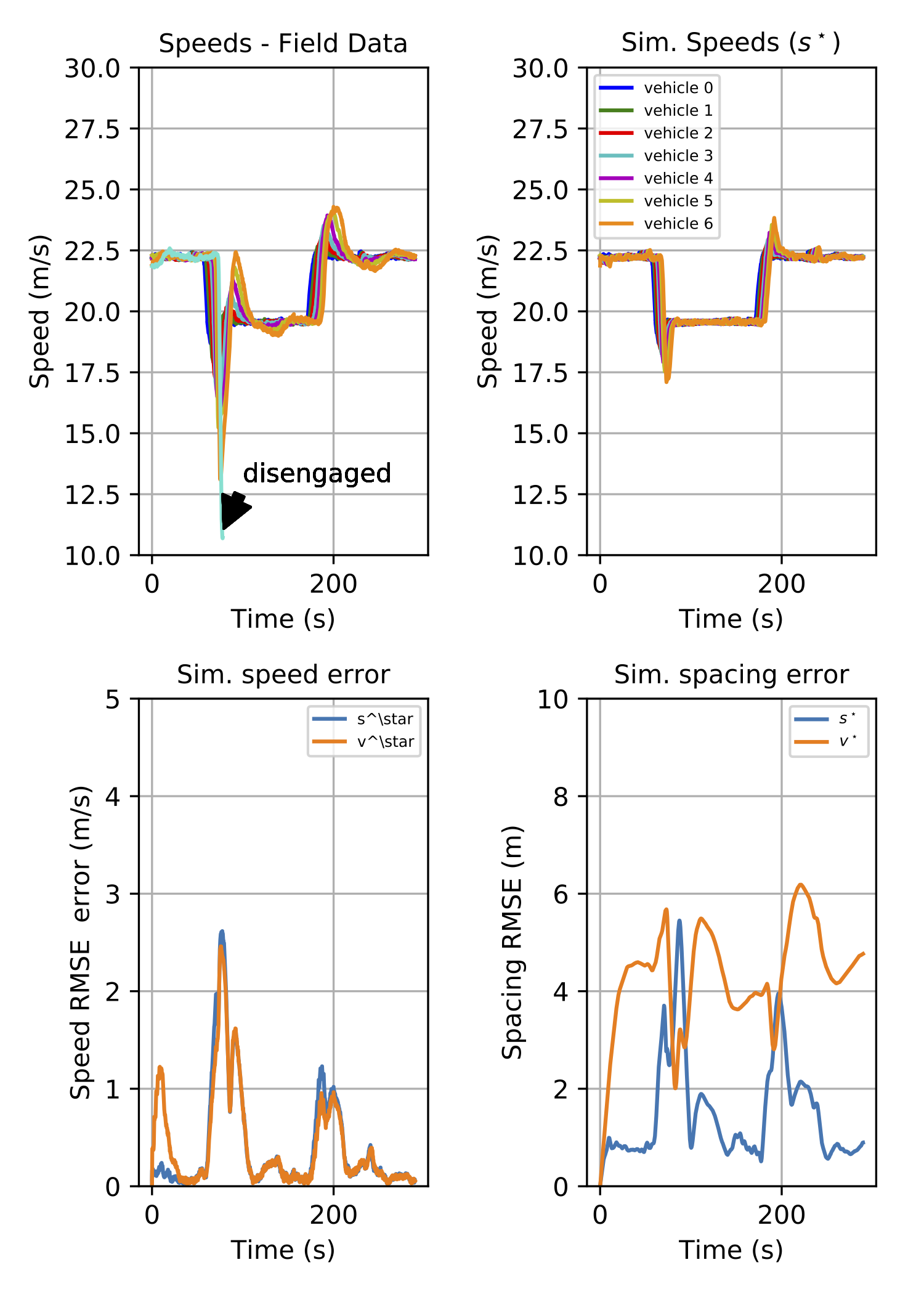}
  \subcaption{Relative speed and position error for the Gipps model.}
  \label{fig:platoongipps}
\end{minipage}
\caption{Validation results for OVRV, IDM, and Gipps models. In each subfigure, the top left shows the observed speeds for the 8 vehicles in the platoon. The top right graph shows the simulated speed using the corresponding car-following model for $s^\star$. The middle left and middle right graph depict, respectively, the speed RMSE and spacing RMSE over time with blue lines representing $s^\star$ and orange lines for $v^\star$.}
\end{figure}

Figure~\ref{fig:platoonovm} shows the simulation results for the OVRV model. The top left Figure depicts the observed speeds for the vehicle 0 (lead) to 7 (last vehicle in the platoon). Note that as mentioned in~\cite{gunter2019are}, vehicle 7 drove below the minimum operating speed of this particular vehicle's ACC controller and thus disengaged the ACC after the first deceleration. The specific vehicle disengages at speed of 12 m/s. The top right graph depicts the simulated speeds. Since the speed profiles for both objectives A and B (minimum spacing error and speed error, respectively) we depict the speed profile for parameter set $s^\star$. Qualitatively, the simulated model replicates key features of the platoon dynamics. The field data shows driving behind the lead vehicle at a constant speed and then transitioning between two set speeds. This occurs on the simulated model as well as seen in Figure~\ref{fig:platoonovm}. \edit{In addition, the amplitude of oscillations is captured in the simulated trajectories.} The exceptions are were the last two vehicles in the field experiment had to decelerate more suddenly than the previous vehicles.

The bottom left and right graphs of Figure~\ref{fig:platoonovm} show, respectively, the speed and spacing RMSE across the 6 vehicles over time. Observe that after reaching a steady speed on the beginning of the simulation, the solution $v^\star$ presents slightly smaller speed error compared to objective A. On the other hand, solution $s^\star$ achieves significantly smaller spacing error throughout the simulation period.

Figures~\ref{fig:platoonidm}  and~\ref{fig:platoongipps} depict the simulated platoon results for IDM model and Gipps model, respectively. The results for IDM is similar to OVRV model. All the models reproduce the amplification of perturbations in the platoon with following vehicles reaching minimum speeds after deceleration and higher speeds after the lead vehicles recover the speed of 22.4 m/s. However, observe that both OVRV and IDM models reproduces an oscillatory behavior when reaching the equilibrium speeds which Gipps model does not. Observe that following the first deceleration, the last vehicle in the platoon reach a speed of around 22.5 m/s when catching up the lead vehicles after overreacting when decelerating. This behavior does not occur for Gipps model. Instead, vehicles monotonically reaches the equilibrium speeds.

\begin{figure}
    \centering
    \begin{minipage}{0.45\textwidth}
  \includegraphics[width=\textwidth]{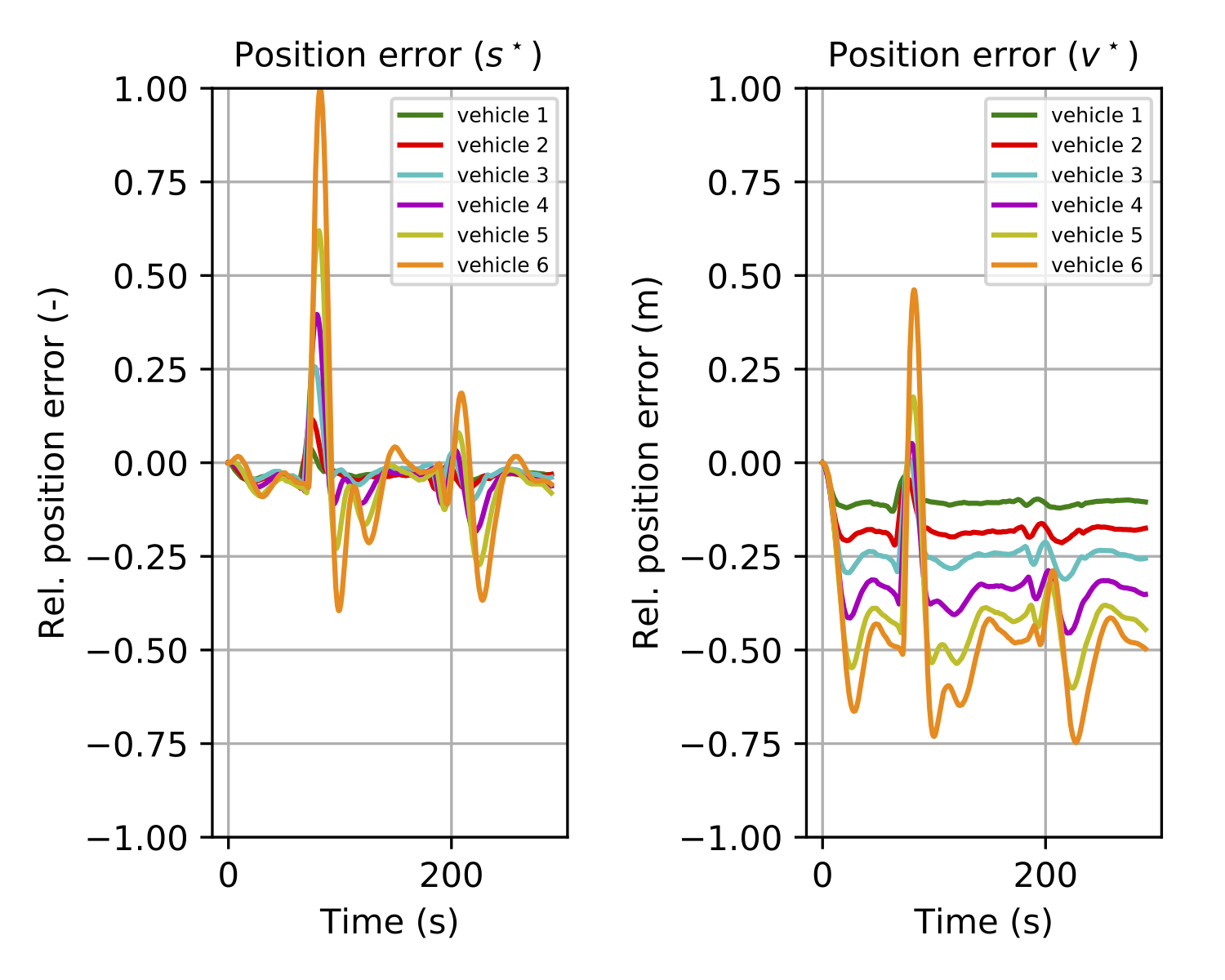}
  \subcaption{Relative speed and position error for the OVRV model.}
  \label{fig:relovm}
\end{minipage}
\qquad
\begin{minipage}{0.45\textwidth}
\includegraphics[width=\textwidth]{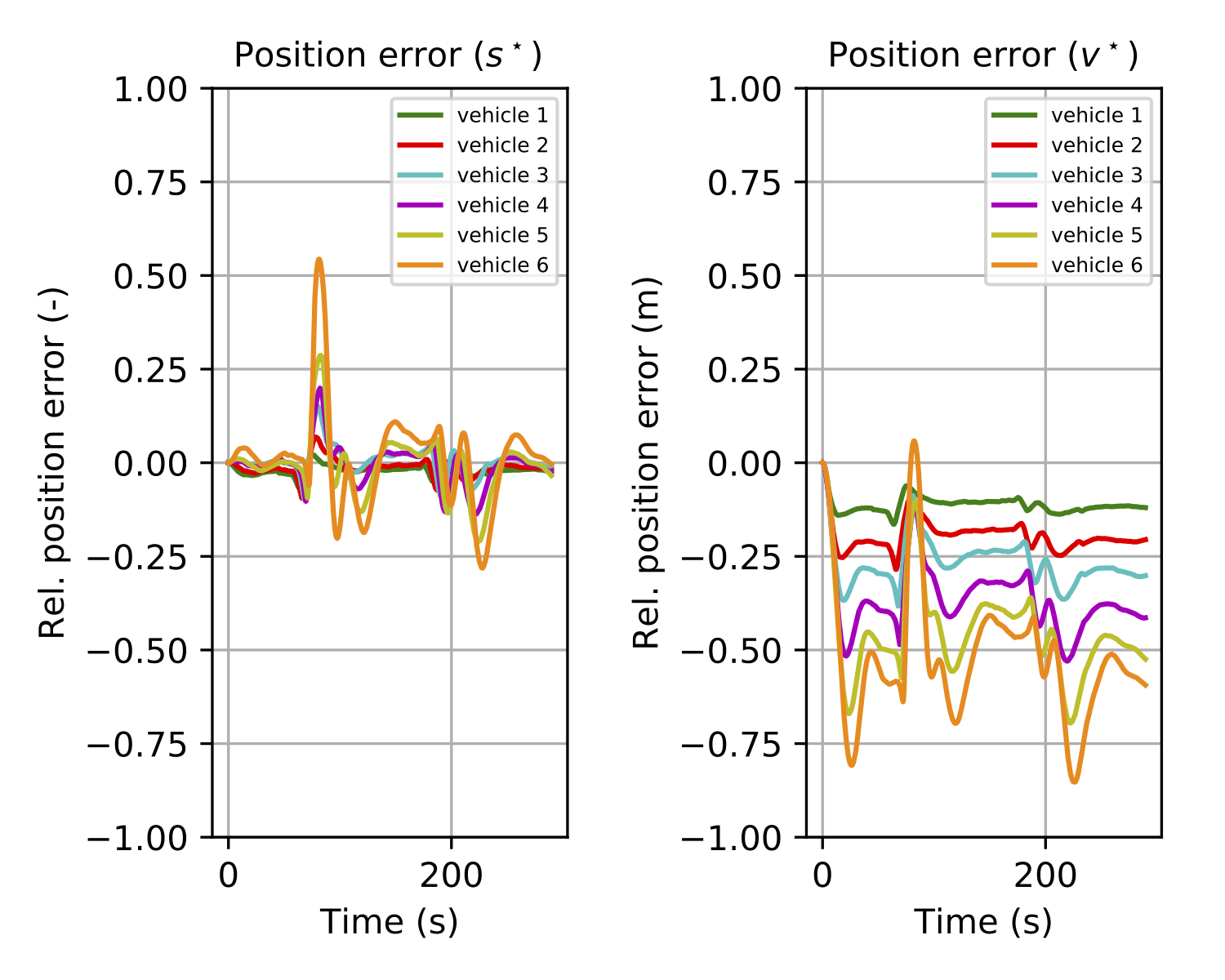}
  \subcaption{Relative speed and position error for the IDM model.}
  \label{fig:relidm}
\end{minipage}
\qquad
\begin{minipage}{0.45\textwidth}
  \includegraphics[width=\textwidth]{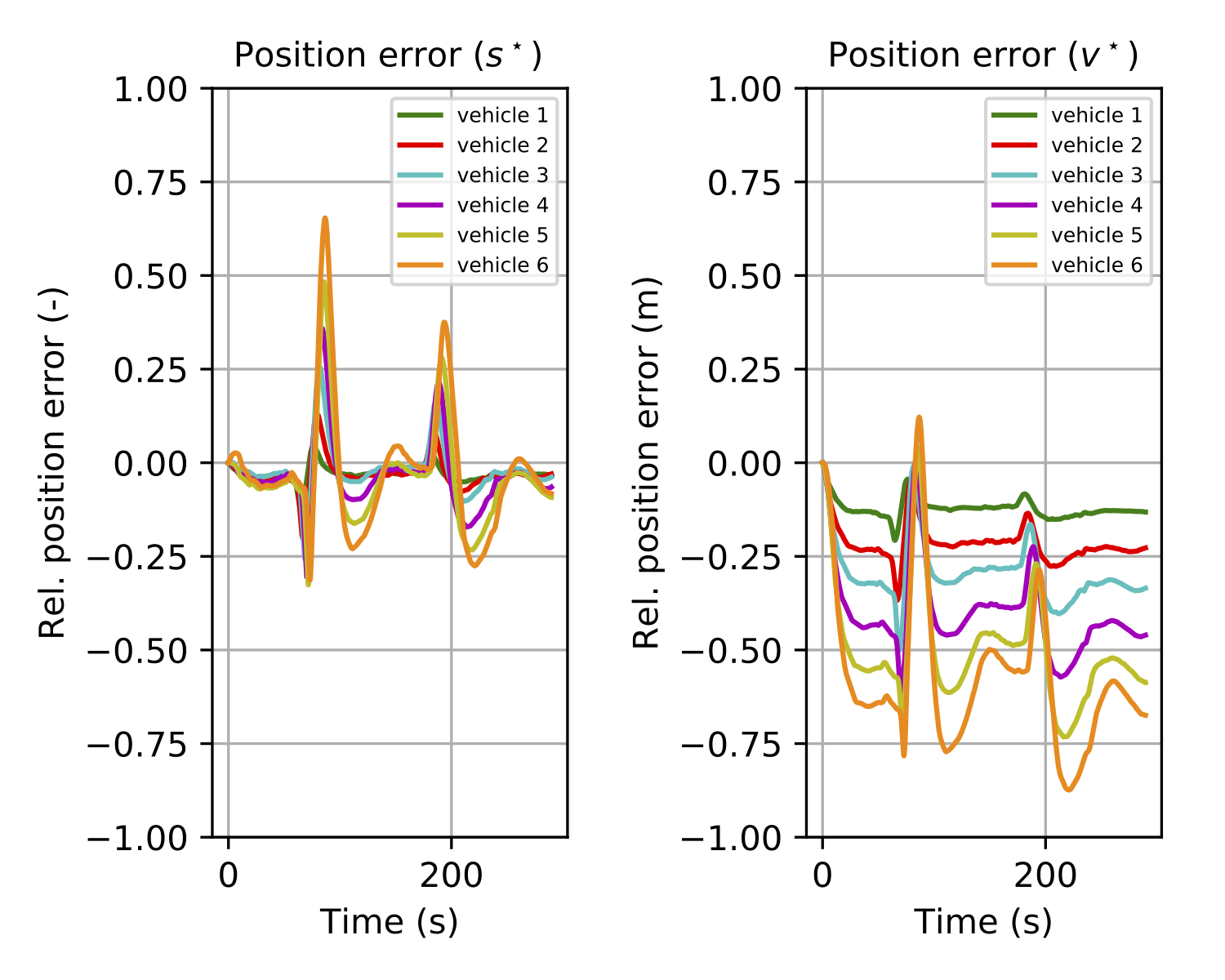}
  \subcaption{Relative speed and position error for the Gipps model.}
  \label{fig:relgipps}
\end{minipage}
\caption{Relative speed and position errors for the platoon validation dataset.}
\end{figure}

We show in Figures \ref{fig:relovm}, \ref{fig:relidm}, and \ref{fig:relgipps} show the difference between the observed and simulated position for all different vehicles for OVRV, IDM, and Gipps models respectively. All values are normalized so that 1 refers to the maximum position error observed across all experiments. Solution $s^\star$ is depicted at the left and solution $v^\star$ at right. For the three models, the results for solution $v^\star$ present a spacing error in steady state of (as shown in the bottom right graph of Figures~\ref{fig:platoonovm}, \ref{fig:platoonidm} and \ref{fig:platoongipps}) that accumulates across the platoon. This leads to an error 5 times larger by the end of the simulation. In all, these results validate the calibration results in the previous section \ref{sec:results} that targeting spacing error is more suitable for calibrating the model parameters.

\section{Conclusions}\label{sec:conclusions}
A calibration approach was proposed for the analysis of field data of the car following behavior of commercially available ACC vehicles under various driving conditions. The use of a multi-objective calibration approach~\edit{applied specifically to data collected from commercially-available ACC vehicles} is the key element that makes this work distinct from the previous literature. The key findings identified from the calibrated models are three-fold: (i) the OVRV and IDM presented comparable performance on the two objectives considered with a slight advantage to OVRV whereas the Gipps model under-performed the other models\edit{, likely because the OVRV is based on the commonly assumed constant time gap control law}; (ii) the trade-off curves in all models, and especially OVRV and IDM, presented a somewhat flat Pareto frontier where reducing spacing errors leads to a small increase in speed error, strongly supporting recent analytical work \cite{punzo2016speed}, which showed spacing error should be prioritized when applying single-objective calibration. The multi-objective approach is especially suitable for this analysis; (iii) the results related to speed-spacing relationship in steady-state are the ones influencing the higher spacing error. The validation of the results in the platoon experiment confirms these findings. \edit{Additionally, this work provides the calibrated parameter values for each vehicle and each model, which can be used for future simulation-based analysis of the traffic flow impacts of ACC vehicles.}

The discontinuity that arises in the Gipps model when switching from free-flow to following mode is a key reason for the model's poor performance when describing commercially-available ACC vehicle dynamics. The observed data do not appear to present such speed discontinuity. However, the IDM and OVRV models do not present this discontinuity. Importantly, this may only be the case for the specific ACC vehicle dynamics, and cannot be transferred to human-driven vehicles. Also, it is important to point out that despite better fit, the OVRV and IDM models cannot reproduce all features observed in the experimental data. In particular, on the transition from low to high speeds the errors increase significantly as can be observed in Figures \ref{fig:timeseriesOVRV} and \ref{fig:timeseriesidm}.

One major source of disturbances on highways that are not considered in the current modeling framework are lane changes. Therefore, we plan to continue this work by integrating car-following models and lane-changing models to create a comprehensive model for ACC vehicle behavior in realistic traffic flow. This will allow us to analyze the true impact of commercially available ACC vehicles on the overall traffic flow at different market penetration rates. \edit{In addition, we are interested in including other potential metrics in the multi-objective approach, especially including quantitative metrics related to the speed and amplitude of oscillations such as the ones proposed in \cite{zheng2011freeway}.}

\section{Acknowledgements}
This work is supported by the Faculty Fellows Program at the Center for Transportation Studies at the University of Minnesota and by the U.S. Department of Energy Vehicle Technologies Office under the Systems and Modeling for Accelerated Research in Transportation Mobility Laboratory Consortium, an initiative of the Energy Efficient Mobility Systems Program.

\bibliographystyle{unsrt}
\bibliography{refs}

\end{document}